# Crosstalk between non-processive myosin motors mediated by the actin filament elasticity

Oded Farago[1] and Anne Bernheim-Groswasser[2]

[1]Department of Biomedical Engineering, [2]Department of Chemical Engineering, Ben-Gurion University of the Negev, Beer-Sheva 84105, ISRAEL.

**Abstract**

Many biological processes involve the action of molecular motors that interact with the cell cytoskeleton. Some processes, such as the transport of cargoes is achieved mainly by the action of individual motors. Other, such as cell motility and division, require the cooperative work of many motors. Collective motor dynamics can be quite complex and unexpected. One beautiful example is the bidirectional ("back and forth") motion of filaments which is induced when the motors within a group exert forces in opposite directions. This review tackles the puzzle emerging from a recent experimental work in which it has been shown that the characteristic reversal times of the bidirectional motion are practically independent of the number of motors. This result is in a striking contradiction with existing theoretical models that predict an exponential growth of the reversal times with the size of the system. We argue that the solution to this puzzle may be the crosstalk between the motors which is mediated by the elastic tensile stress that develops in the cytoskeleton track. The crosstalk does not directly correlate the attachment and detachment of the motors, which work independently of each other. However, it highly accelerates their detachments by making the detachment rates system size dependent.

## 1. Bidirectionl motion as a tug-of-war between motors

Cells utilize biological motors for active transport of cargo along their respective filaments to specific destinations (1). Various types of motor proteins have different preferred directions of motion. Most kinesins and myosins, for instance, move towards the plus end of microtubules (MTs) and actin filaments, respectively (2). Others, such as Ncd and myosin VI, move towards the minus end (3,4). While some processes, such as the transport of cargoes is achieved mainly by the action of individual motors, other processes, such as cell motility and mitosis, require the cooperative work of many motors. Muscle contraction, for instance, involves the simultaneous action of hundreds of myosin II motors pulling on attached actin filaments and causing them to slide against each other (5). Similarly, groups of myosin II motors are responsible for the contraction of the contractile ring during cytokinesis (6). In certain biological systems, cooperative behavior of molecular motors produces oscillatory motion. In some insects, for instance, autonomous oscillations are generated within the flight muscle (7). Spontaneous oscillations have also been observed in single myofibrils *in vitro* (8). Finally, dynein motors could be responsible for the oscillatory motion of axonemal cilia and flagella (9,10).

One of the more interesting outcomes of cooperative action of molecular motors is their ability to induce bidirectional motion. ``Back and forth" dynamic has been observed in various *in-vitro* motility assays including: (i) NK11 (kinesin related Ncd mutants which individually exhibit random motion with no preferred directionality) moving on MTs (11), (ii) mixed population of plus-end (kinesin-5 KLP61F) and minus-end (Ncd) driven motors acting on MTs (12), and (iii) myosin II motors walking on actin filaments in the presence of externally applied forces (13,14). In cells, the motion of cargos along MTs is often bidirectional, which is attributed to the presence of opposing kniesin (plus-end directed) and dynein (minus-end directed) motors (15-17). Early studies of this phenomenon suggested that there exists a "coordinated switching" mechanism, regulated by proteins, which allows only one type of motors to bind to the cargo at a given time. Thus, the kinesin motors are turned off when the dynein motors are pulling the cargo and vice versa, and they do not actually work against each other. An alternative mechanism has been recently proposed in which the motors are engaged in a *tug-of-war* (TOW) on the cargo. In the TOW model, the cargo moves in the direction of the motor party that exerts the larger force. The balance of power is shifting between the two parties as a result of stochastic events of binding and unbinding of motors. The main feature of the TOW model lies in the fact that the unbinding rates depend exponentially on the force load experience by the motors, which itself depends on the number of attached plus-directed and minus-directed motors. This leads to a very rich dynamic behavior which is very sensitive to the model parameters (which include the stall force, detachment force, unbinding and binding rates, forward velocity, and superstall velocity amplitude – overall 6 parameters for each motor type) (18). Specifically, for certain sets of parameter values, the motion is bidirectional, i.e., switches between periods of plus-directed and minus-directed movements. Interestingly, during these periods of unidirectional motion, the motors that win the contest cause the detachment of all the motors of the other type. This pattern of bidirectional motion had been erroneously associated with the coordinated transport mechanism. Recent experiments, which have carefully analyzed the bidirectional transport of vesicles along MTs, concluded that the dynamics is indeed consistent with the tag-of-war mechanism (19,20).

Bidirectional motion does not necessarily require the existence of two type of motors, but may be also observed when one group of motors is driving the motion of *filaments and bundles* *with mixed polarities* (21,22). Muscles and stress fibers, for instance, consist of anti-parallel actin filaments with partial overlap. When myosin motors operate on such structures, they cause these filaments to move in opposite directions and contract. Recently, we reported on a novel motility assay in which we generated actin bundles consisting of filaments with alternating polarities (22). The filaments with alternating polarities are formed by short polar actin segments which are transported by the motors, brought to close proximity with each other, and then fused, presumably by motors which were left in the solution but do not reside on the surface. When placed on a bed of immobilized myosin II motors, these mixed polarity bundles exhibit bidirectional motion because the motors that act on the different polar segments apply forces which are opposite in their directions. The competition between the motors working in opposite directions and the resulting bidirectional motion can be analyzed within the framework of the TOW model. However, one must bear in mind that the TOW model has been originally developed to describe the transport of vesicles by a relatively small number of motors (typically $N \leq 10$), while in ref. (22) the dynamics usually involve several hundreds and even thousands of motors ($N \geq 500$). Hexner and Kafri (23) have recently analyzed the TOW model in the large $N$ limit and found two patterns of bidirectional motion: The first one is of a rapid oscillating-like motion, with microscopic reversal times of the order of the ATPase cycle (i.e., the typical attachment time of a single motor to the filament). The second one is bidirectional motion with macroscopically large reversal times that grow exponentially with $N$. For sufficiently large $N$ ($N \geq 1000$), one effectively reaches the ''thermodynamic limit'' in which the reversal time diverges and motion persists in the direction chosen randomly at the initial time. This unidirectional motion is one of the two possible steady-state solutions of the dynamics. Moving in the opposite direction is the other steady-state solution, and the bidirectional motion represents the occasional transition between these two states. For large $N$, the transition probability between the two states vanishes, in a manner which is analogous to the spontaneous symmetry breaking in ferromagnetic materials below the critical temperature and in the absence of external magnetic field.

By using a-polar actin bundles of different sizes, we were able to measure the dependence of the reversal times $\tau_{rev}$ on $N$ (22). The experimental results, which are summarized in Fig. 1, show that while $N$ varies over half an order of magnitude, the corresponding $\tau_{rev}$ are similar to each other ($3 < \tau_{rev} < 10$ s) and show no apparent correlation with $N$. These results are not well described by neither the "rapid-oscillations" nor "the bidirectional motion with exponentially diverging reversal times" scenarios of the TOW model at large $N$ (23). On the one hand, the experimentally measured reversal times were of the order of several seconds, i.e., 2-3 orders of magnitude larger than the typical ATPase cycle of individual myosin II motors. On the other hand, there was no apparent correlation between $N$ and $\tau_{rev}$, and certainly no exponential dependence.

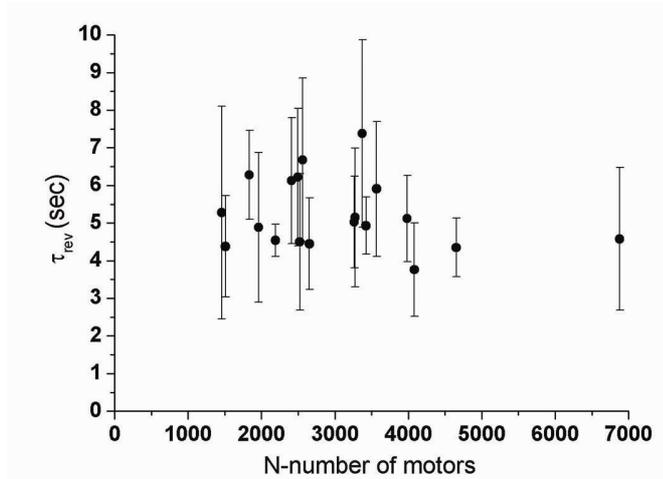

**Figure 1: The characteristic reversal time of 19 different bundles as a function of the number of working motors.**

How can we explain these experimental results? To this end, let us first understand the origin of the exponential dependence of $\tau_{rev}$ on $N$. Consider an actin track with alternating polarities which in the motility assay interacts with $N$ motors. The motors can be divided into three groups: (i) those which are disconnected from the track and do not apply any force, (ii) those which are connected to segments whose plus end points to the right and, therefore, push the actin track to the left, and (iii) those which are connected to segments whose plus end points to the left and push the actin track to the right. The last two groups compete with each other, and the occasional reversals of the transport direction reflect the "victory" of one group over the other during the respective time intervals. The balance of power is shifting between the two motor parties as a result of stochastic events of binding and unbinding of motors to the cytoskeletal track. One can write a set of coupled master equations that describes the transitions between these three groups of motors (23,24). These equations have two identical, except for sign reversal, steady-state solutions corresponding to right and left motion. In each of these solutions, the number of motors working in the direction of motion, $\langle N_+ \rangle = p_+ N$, is larger than in the opposite direction, $\langle N_- \rangle = p_- N$. In general, $p_+$ and $p_-$ depend on the biochemical features of the motors and the track and on the ATP concentration. In the TOW model, they also depend on $N$, through the intricate coupling between $N_+$ and $N_-$ via the force balance equation, the force-velocity relation, and the load dependence of the unbinding rates (18). However, as carefully analyzed in ref. (23), the dependence of $p_+$ and $p_-$ on the system size disappears in the limit of a large number of motors. In fact, in the TOW model, one usually finds that $p_- \ll p_+$ since the motors that lose the TOW contest tend to have a much larger unbinding rate than those that win. A change in directionality, namely a switch from one steady state solution to the opposite one, will occur only if almost all the winning motors will be detached from the track simultaneously (more precisely- within a short microscopic time interval $\tau_0$). This becomes a very rare event in the thermodynamic limit. If the motors act independently of each other, the occurrence probability of such an event drops exponentially with $N$, and the typical reversal time $\tau_{rev}$ (which is inversely proportional to the occurrence probability) grows exponentially with $N$

$$\tau_{rev} \sim \tau_0 \exp(cN), \qquad (1)$$

where $c$ is a dimensionless constant. In a more detailed calculation (24), we showed that

$$c = -\ln\left[1 - \left(\sqrt{p_+} - \sqrt{p_-}\right)^2\right]. \qquad (2)$$

The exponential growth of the reversal times with the number of working motors is a characteristic footprint of cooperative motor dynamics. One may consider $\tau_{rev}$ as a measure for the degree of cooperativity between the motors. The more cooperative the motors are, the more persistent is the movement and the longer are the periods of unidirectional transport. As demonstrated by the above argument leading to eqns. (1) and (2), this behavior is related to the lack of correlations between the detachment events of different motors. Our experimental results showing that $\tau_{rev}$ does not grow exponentially with $N$ suggest the existence of some coupling between the motors. A possible origin for this coupling is the elasticity of the actin track which may mediate crosstalk between the motors. The elasticity-mediated crosstalk between the motors can be manifested in to two possible ways: (i) The attachments and detachments of the different motors may become correlated, and (ii) the attachments and detachments of the different motors remain uncorrelated (i.e., each motor binds to or unbinds from the track independently of the binding state of the other motors), but the attachment and detachment *rates* of each motor depend on the binding states of the others. Below, we argue that the latter effect provides a reasonably adequate explanation for the experimental results. The analysis of this effect is done using a ratchet model – a theoretical framework which is frequently used for studying motor protein systems.

**2. Ratchet models for motor systems**

Brownian ratchet theory refers to the phenomenon of motion induced by non-equilibrium fluctuations in an isothermal medium with broken spatial symmetry. This concept was first introduced by Smoluchowski (25), and later revisited by Feynman (26). In the 1990s, interest in the ratchet mechanism has been revived as a possible explanation for transport phenomena of molecular motors (27-31). Here, we give a brief account on the topic, with emphasis on ratchet models for collective motor dynamics. For a very detailed review which summarizes the development of the field, see ref. 32.

The ratchet model assumes that motor molecules are Brownian particles that move in a locally asymmetric periodic potential. This potential represents the binding energy, $U_{attached}(x)$, between the filament and the motor. Its periodic asymmetric form reflects the periodic nature of the filament and its polarity. A motor which is not connected to the track is considered to be in a higher energy state where it experiences a uniform potential, $U_{detached}(x) = const$. At thermodynamics equilibrium, the transition probabilities between the attached and detached states obey detailed balance:

$$p_{attached}(x) / p_{detached}(x) = \exp\left[-\left(U_{attached} - U_{detached}\right)/k_B T\right]. \qquad (3)$$

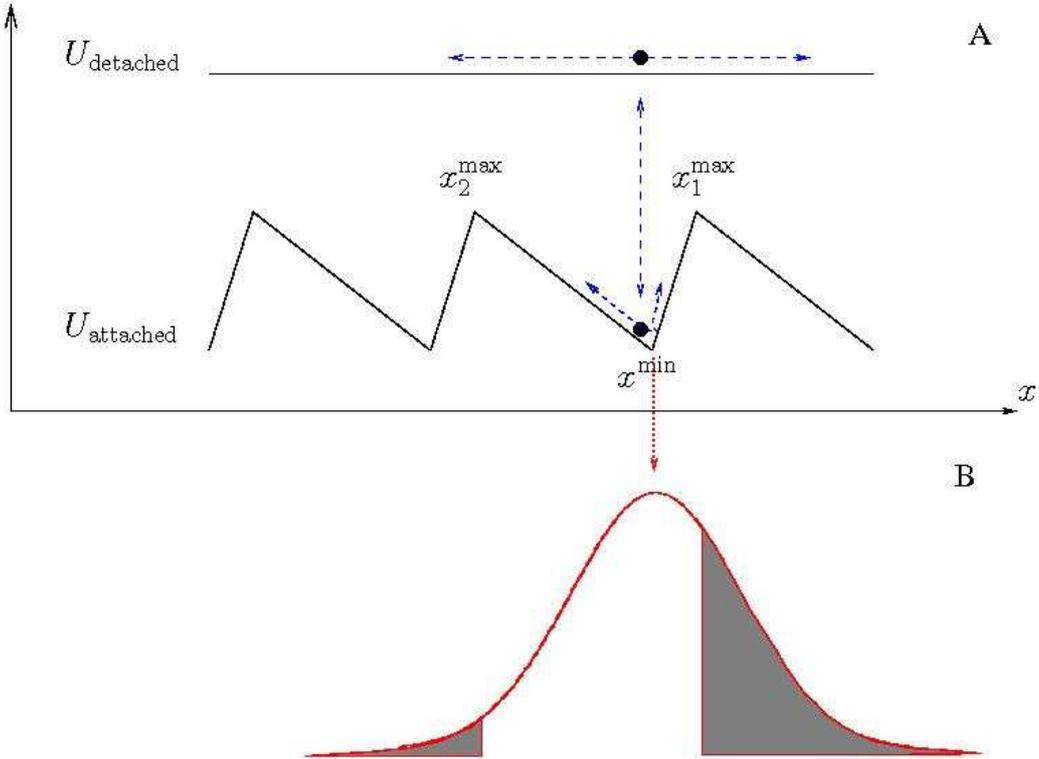

**Figure 2: The working principle of an "on-off" Brownian ratchet.** (A) In the attached ("on") state the particle is localized in the potential minimum at $x^{\min}$. Once going to the detached ("off") state, its distribution function spreads. When the particle switches back to the on state, the particle is captured at the same basin of attraction or in the adjacent potential wells, depending on the distance it travelled from $x^{\min}$. (B) If the partivle switches periodically between the attached and detached states (i.e., spends a fixed amount of time in the detached state), the probabilities to move to the neighbor unit cells are given by the grey shaded areas under the Gaussian distribution function. Because of the asymmetric shape of the potential, the probability to move to the right is larger than the probability to move to the left, which result in a drift of the particle in the right direction.

An illustration of the system is given in Fig. 2A showing a particle diffusing along an asymmetric saw-tooth ratchet potential. The particle will spend most of its time in the lower energy attached state, and within this state it is likely to be found near the minimum of the potential well at $x^{\min}$. Occasionally, the particle will be thermally excited to the detached state in which it does not interact with the potential and, therefore, can diffuse freely. If the particle remains long enough in this state, it can diffuse across the spatial position of the maximum of the ratchet potential located at $x_1^{\max}$ on the right and at $x_2^{\max}$ on the left. If that happens, then once the particle gets back into the attached ground state it will be captured in the adjacent potential well. A detailed calculation (see ref. 32 for a discussion on this non-intuitive conclusion) shows that in spite of the broken spatial symmetry of the ratchet potential (namely, the fact that $\left|x_1^{\max} - x^{\min}\right| \neq \left|x_2^{\max} - x^{\min}\right|$), the particle's motion will not be biased toward one of the directions. This somewhat surprising result is a direct consequence of the second law of thermodynamics that prohibits the conversion of heat into work and motion in an isothermal system.

If the system is driven out of equilibrium (which in motor systems occurs via the continuous input of ATP chemical energy), the conditions of detailed balance would be violated. In such a case, the particle can harness the Brownian thermal noise to rectify the diffusion and drift, as demonstrated in the following example: Suppose that the transition rules between the states are changed and instead of eqn. (3), the particle oscillates periodically between the attached and detached states. Under this non-equilibrium transition rule, the particle spends longer times in the detached state compared to equilibrium conditions. In the attached state, the particle distribution function is concentrated at $x^{\min}$. Once going to the excited detached state, the distribution function spreads out to a symmetric Gaussian distribution (see Fig. 2B). The grey shaded areas under this Gaussian distribution give the probabilities that the particle diffuses beyond $x_1^{\max}$ and $x_2^{\max}$ - events that would place the particle in the adjacent well once it falls back to the attached state. As a result of the broken symmetry of the ratchet potential, the traveling probabilities to the left and right are not equal and the particle is expected to move preferentially in one of the directions (to the right in the case depicted in Fig. 2).

A ratchet model for motor molecules that cooperate in large groups has been firstly introduced in 1995 by Jülicher and Prost (JP) (33,34). As in ratchet models for a single motor, the motors in the JP model are represented by particles that move along a periodic potential. However, instead of moving individually, they move in cooperation as if they are connected to a rigid rod. The group velocity is determined by the total force exerted on the motors by the ratchet potential. This force is the sum of forces experienced by the motors which may be (i) either positive of negative, depending on the coordinate $x$ of the attached motor along the ratchet potential, or (ii) vanish, in case the motor detaches from ratchet potential. In accordance with the second law of thermodynamics, the motor will not exhibit spontaneous motion if the transition rates between the attached and detached states satisfy detailed balance. Directed motion is possible only if both detailed balance and the symmetry of the potential are broken.

When the ratchet potential is symmetric, there is obviously no preference to any of the directions. The cooperative motion exhibited by the motors is bidirectional. In this case, the ratchet model predicts the same two scenarios found by the TOW model for equal numbers of motors working in the opposite directions: If the deviation from detailed balance is small (which corresponds to low concentrations of ATP), the motion is characterized by rapid fluctuations between left and right movements (33). Rapid oscillations are also expected in systems with a small number of motors. Above a critical value of the non-Brownian noise, the system undergoes a spontaneous symmetry breaking and moves in one of the two possible directions. The probability to reverse the motion diminishes exponentially with the size of the system, which is equivalent to saying that the characteristic reversal time grows exponentially with number of motors. This observation is a direct consequence of the absence of correlations between the different motors – a feature shared by the TOW and the ratchet models in the thermodynamic limit.

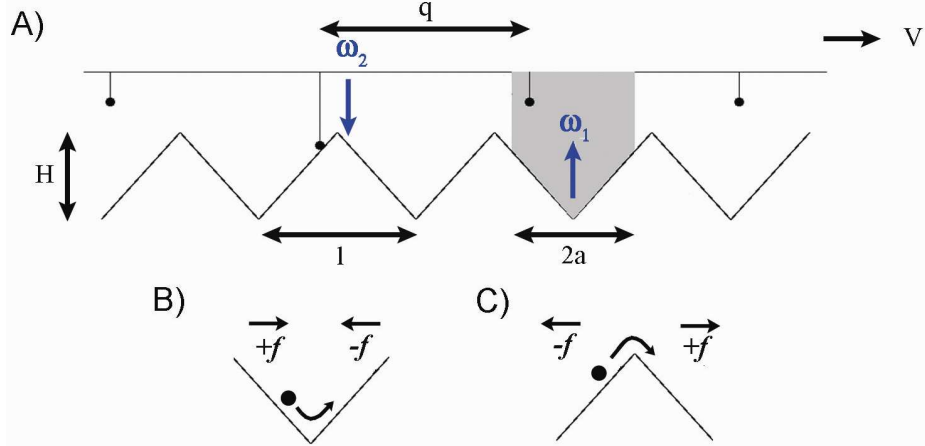

**Figure 3:** (A) $N$ point particles (representing the motors) are connected to a rigid rod with equal spacing $q$. The particles move in a periodic, symmetric, saw-tooth potential with period $l$ and height $H$. The particles experience the force of the potential only in the at attached state. The detachment rate from the potential $\omega_1$ is localized in the shaded area of length $2a < l$, while the attachment rate $\omega_2$ is located outside of this region. (B) When an attached motor crosses a minimum of the periodic potential, the direction of the force that it experiences changes from positive (in the direction of motion) to negative (against the direction of motion). (C) When an attached motor crosses a maximum of the periodic potential, the direction of the force that it experiences changes from negative to positive.

Fig. 3A presents a model with a symmetric ratchet potential, which is a modified version of the model originally introduced in 2002 by Badoual, Jülicher and Prost (35). We consider the 1D motion of a group of $N$ point particles connected to a rigid rod with equal spacing $q$. The particles move in a periodic, symmetric, saw-tooth potential, $U(x)$, with period $l$ and height $H$. The model requires $q$ being larger than and incommensurate with $l$ (which is indeed the case in the motility assay where $l = 5$ nm, while the density of the motors on the surface is typically such that $q = 6 - 7$ nm). The instantaneous force between the track and the motors is given by the sum of all the forces acting on the individual motors:

$$F_{tot}(t) = \sum_{i=1}^{N} f_i^{motor} = \sum_{i=1}^{N} \left[ -\frac{\partial U(x_1 + (i-1)q)}{\partial x} \right] \cdot C_i(t), \quad (4)$$

where $x_i = x_1 + (i-1)q$ is the coordinate of the $i$-th motor. The function $C_i(t)$ takes two possible values, 0 or 1, depending on whether the motor $i$ is detached from or attached to (0 - detached; 1 - attached) the ratchet potential at time $t$. The group velocity of the motors (relative to the track) is determined by the equation of motion for overdamped dynamics: $v(t) = F_{tot}(t)/\lambda$, where $\lambda$ is the friction coefficient.

The transitions between the states are governed by the following non-equilibrium rules (we ignore the additional equilibrium contribution which is assumed to be much smaller). The motors change their states independently of each other. We define an interval of size $2a < l$ centered around the potential minima (the shaded area in Fig. 3A). If located in one of these regions, an attached motor may become detached (1→0) with a probability per unit time $\omega_1$. Conversely, a detached motor may attach to the track (0→1) with transition rate $\omega_2$, only if located outside this region of size $2a$. These non-equilibrium transition rules ensure that the motion in a given direction is persistent, for the following reason. Suppose that there are more motors attached on the left side of the minimum than on the right side. The former experience positive forces, while the latter are subjected to negative forces. Since the total force is positive, the collective motion of the motors is in the positive direction, i.e., to the right. When an attached motor crosses one of the minima or maxima, the direction of the force it experiences flips. Crossing a minimum will increase (decrease) the number of motors experiencing a force opposite to (in) the direction of motion (Fig. 3B). Crossing a maximum will have the opposite impact, i.e., the number of forces in (opposite to) the direction of motion will increase (decrease) by one (Fig. 3C). By allowing the motors to detach from the potential around the minimum and attach around the maximum, we ensure that the scenario depicted in C occurs more frequently than the one in B. In other words, the number of negative forces turning to positive is larger than the number of positive forces flipping to negative. This preserves the bias between the positive and negative forces, which would maintain the direction of the motion. In the thermodynamic limit $N \gg 1$, one can analytically calculate the steady state attachment probability, $p_{att}$, of a motor as a function of the spatial coordinate $-l/2 \leq x \leq l/2$ within the unit cell of the periodic potential (24). Several solutions are plotted in Fig. 4 for positive group velocity $v > 0$, $2a = 0.75l$, and $(\omega_1, \omega_2) = (v/l, v/l)$ (thin solid line), $(4v/l, 4v/l)$ (dashed line), and $(20v/l, 20v/l)$ (thick solid line). Since these solutions correspond to the case that the motors move the right, it is easy to understand why $p_{att}$ reaches its maximum at $x = -a$ (just before the motors enter, from the left, into the central gray-shaded detachment interval, $-a < x < a$) and its minimum at $x = a$ (just before leaving the central detachment interval through the right side). We also notice that when the off rate $\omega_1 \gg v/l$, $p_{att}$ drops very rapidly to near zero in the detachment interval. When the attachment rate $\omega_2 \gg v/l$, $p_{att}$ increases exponentially fast for $x > a$ and rapidly reaches the maximum possible value $p_{att} = 1/l$. Overall, the total attachment probability on the left side of the minimum ($-l/2 \leq x \leq 0$) is larger than the total attachment probability on the right side ($0 \leq x \leq l/2$) which reflects the tendency of the system to propagate to the right. If one assumes that the system propagates to the left ($v < 0$), the other steady state solution is obtained, which is simply a mirror reflection of the first solution with respect to $x = 0$.

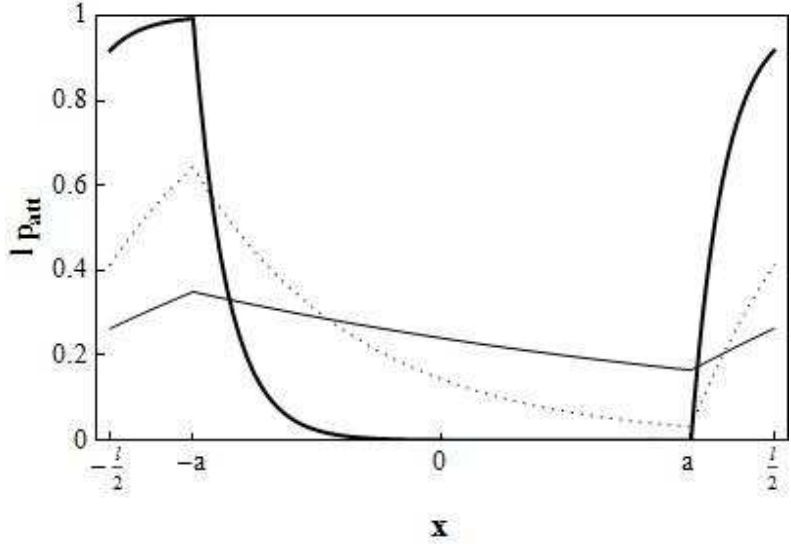

**Figure 4: The steady-state probability density, $p_{att}$, as a function of $x$, the position within a unit cell of the periodic potential. The functions plotted in the figure correspond to $2a = 0.75l$, and $(\omega_1, \omega_2) = (v/l, v/l)$ (thin solid line), $(\omega_1, \omega_2) = (4v/l, 4v/l)$ (dashed solid line), and $(\omega_1, \omega_2) = (20v/l, 20v/l)$ (thick solid line).**

The fact that the tracks in the motility assay consist of polar segments with randomly alternating polarities can be incorporated into the model by introducing an additional force $f_{ran}$ (denoted by red arrows in Fig. 5A) in each period of the ratchet potential. Each force points either to the right or left depending on the local polarity of the corresponding segment, and the sequence of forces represents a given "realization" of a track with alternating polarities. Their sum determines the net polarity of the track. If it vanishes, the track is globally a-polar, and from symmetry considerations, the dynamics of the motors should be bidirectional with no net drift. The additional random forces cause a reduction in the characteristic reversal times of the bidirectional motion. However, as long as $f_{ran}$ is smaller than the slope of the symmetric saw-tooth potential, the reversal times would still grow exponentially with the $N$, as demonstrated by the open circles in Fig. 5B showing the average $\tau_{rev}$ computed for 40 different realizations of random, overall a-polar, tracks. When $f_{ran}$ becomes larger than the slope of the ratchet potential, the motion becomes rapidly oscillating with microscopically small $\tau_{rev}$. A closely related situation, namely the transition from bidirectional to rapidly oscillating dynamics upon increasing the magnitude of a local random field, has been found in the TOW model with spatial disorder resulting from the inhomogeneous distribution of motors on the surface (23).

**3. Cooperative dynamics on elastic tracks.**

The motors in our model do not interact with each other and have no mutual influence on the each other's state (attached/detached). As discussed in section 1, the lack of crosstalk between motors leads to the computationally observed exponential dependence of $\tau_{rev}$ on $N$. The experimental data, which does not exhibit this exponential dependence, suggest that the "no crosstalk" assumption may not be justified. Crosstalk between motors can be manifested in two different ways: The first option is that the motors interact with each other in a manner that leads to correlations between their states. The correlations may be positive (distinct motors tend to attach and detach together) or negative (attachment of one motor leads to the detachment of the other). The second option is that the motors do not *directly* influence the states of each other, but instead experience a different kind of cooperativity effect that changes their transition rates between states. Below, we demonstrate how the elasticity of the cytoskeleton track mediates this type of effect.

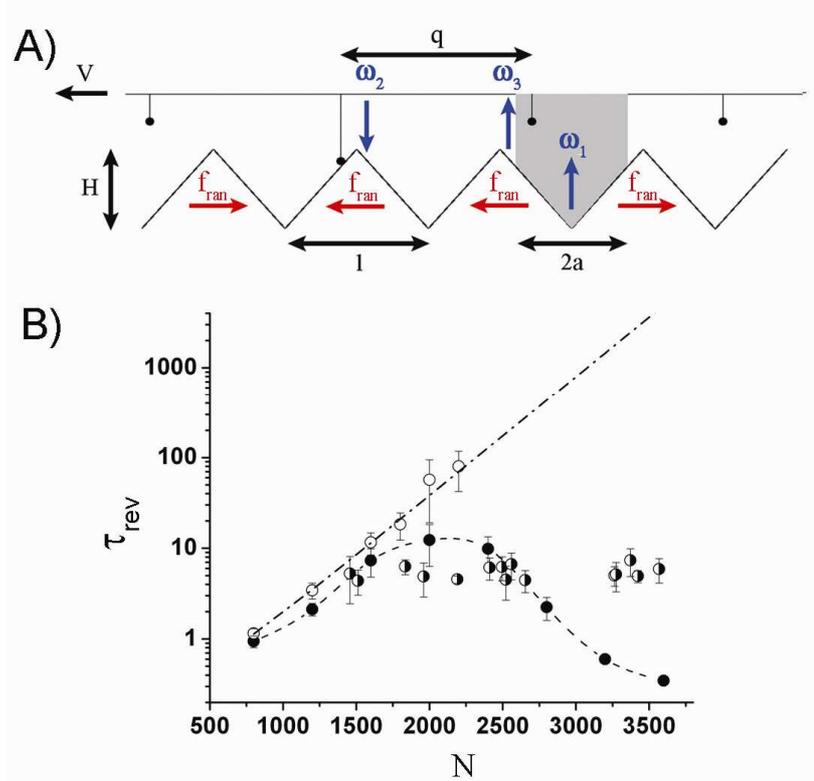

**Figure 5: A ratchet model for cooperative motion of myosin II motors on *elastic* actin tracks *with randomly alternating polarities*.** The model is similar to the one presented in Fig. 3A, with the following additional features: (i) In each periodic unit, there is a random force of size $f_{ran}$, pointing either to the right or to the left (red arrows). (ii) An additional off rate $\omega_3$, permitted outside the grey shaded area, is introduced. The model parameters representative of myosin II-actin systems which were used in the simulations are discussed in ref. 22. (B) The mean reversal time, $\tau_{rev}$, as a function of the number of motors $N$, computed for 40 different realizations. The error bars represent the standard deviation of $\tau_{rev}$ between realizations. Open circles denote the results for a rigid track [$\alpha = 0$ in eqn. (6)], while the solid circles correspond to the case of an elastic track with $\alpha = 0.0018$ (with the dashed line serving as a guide to the eye). The half-filled circles denote the experimental results, also presented in Fig. 1.

Generally speaking, the rates of transitions between the states depend on many biochemical parameters, most notably the types of motors and tracks, and the concentration of chemical fuel (e.g., ATP). They may also be affected by the forces induced between the motors and the filament, which result in increase in the configurational energy of the attached myosin motors (36-39) and in the elastic energy stored in the S2 domains of the mini-filaments. The change in the detachment rates of the motors resulting from the loads which they experience is the key element in the TOW model. But in the TOW model the motors act on a rigid cargo. When motors work on an elastic filament, the forces generated in the cytoskeleton track will modify the loads experienced by individual motors (40). However, because the maximum load on each motor is limited (18), this will lead to a renormalization of the mean detachment rate of the motors, but not to an elimination of the exponential dependence of $\tau_{rev}$ on $N$. There is, however, another indirect contribution which turns out to have a very dramatic effect. The binding and unbinding of motors change the elastic energy stored in the elastic track, and the resulting changes in this energy must also be taken into account when the detachment/attachment rates are calculated. As the following scaling argument will show, because of the cooperative nature of the force generation, the detachment of even a single motor may lead to the release of a very significant amount of elastic energy *in the track*. This, and not the load on individual motors, becomes the main factor determining the detachment rates of the motors for large $N$. The situation is analogous to a thermodynamic system with medium-mediated interactions, e.g., the depletion forces between large colloids in polymer solutions whose origin is the change in the configurational entropy of the polymers and not direct interaction between the colloids themselves (41). In contrast to colloids and polymer solutions, in the case of the myosin II motors, the medium through which the motors communicate is *not* the surrounding solvent but the actin. By sensing the changes in the elastic tensile load along the actin, the motors "learn" about the changes in the states of other motors. This information propagates along this elastic cable in the speed of sound and reaches, almost instantaneously, from one side of the actin to a remote motor on the other side. More precisely, a phonon travels a distance of about 10 micro-meters (typical size of a filament in the experiment) in about 10 nano-seconds, which is 5 orders of magnitude smaller than the typical attachment time of a motor (42).

The elastic energy stored in the actin track can be estimated by the following scaling argument: The total elastic energy of the track scales as scales as $E \sim \langle F_{tot}^2 \rangle / k_{sp}$, where $k_{sp}$ is the effective spring constant of the track, and $F_{tot}$ is the total force exerted on the track by $N_c \leq N$ attached motors. The force $F_{tot}$ is the sum of $N_c$ forces working in randomly alternating directions, which implies that $\langle F_{tot} \rangle = 0$ and $\langle F_{tot}^2 \rangle \sim N_c$. The spring constant is inversely proportional to the length of the track, i.e., to the size of the system and to the total number of motors $N$. Thus, the mean elastic energy of the filament scales like

$$E / k_B T \sim N N_c, \qquad (5)$$

which means that the detachment of a motor ($N_c \to N_c - 1$) leads, on average, to an energy gain

$$\Delta E / k_B T = -\alpha N, \qquad (6)$$

where $\alpha$ is some dimensionless parameter. Notice that in eqn. (5), the *energy per spring* grows linearly with $N_c$. Motors that work in randomly alternating directions along the filament generate a higher load in the springs compared to the case of polar filaments where all the motors work in the same direction. In the latter case, the energy stored in each spring is essentially independent of $N_c$, provided that the distribution of attached motors along the filament is homogenous (40).

The important point about eqn. (6) is that the energy released by the detachment of one motor grows linearly with the size of the system. This effect is incorporated within the ratchet model, by introducing an additional off rate $\omega_3 = \omega_3^0 \cdot \exp(\alpha N)$ outside the gray shaded area in Fig. 5A. The model now includes three transition rates: (i) $\omega_1$ - representing the probabilities per unit time of a motor to detach from the track after completing a unit step, (ii) $\omega_2$ - the attachment probability, and (iii) $\omega_3 = \omega_3^0 \cdot \exp(\alpha N)$ - the rate of detachment without completing the stepping cycle, caused by the elasticity effect. For actin-myosin systems we set these rates to: $(\omega_1)^{-1} = 0.5$ ms, $(\omega_2)^{-1} = 33$ ms, $(\omega_3^0)^{-1} = 7500$ ms, and $\alpha = 0.0018$ (22). The solid circles in Fig. 5B depict our computational results for this set of parameters. Instead of an exponential behavior, $\tau_{rev}$ now exhibits much weaker variations with $N$. The mean reversal times computed for $800 < N < 3600$ (which largely overlaps with the estimated range of number of motors in our experiments) are found in the range $1 < \tau_{rev} < 12$ - in a very good quantitative agreement with the corresponding range of experimental results (Fig. 1).

The agreement between the computational and experimental results is quite remarkable in view of the extreme simplicity of the ratchet model that we use. One should nevertheless be aware of the following points of disagreement: (i) The computed reversal times show weak, non-monotonic, dependence on $N$ which is not observed experimentally. (ii) The largest computed $\tau_{rev}$ ($\tau_{rev} = 12$ s for $N = 2000$) is slightly larger than the experimentally measured reversal times. (iii) The computational results for $N < 1000$ and $N > 3000$ cannot be directly compared with experimental results since the corresponding reversal times ($\tau_{rev} < 1$ s) fall below the experimental resolution. The decrease of the computed reversal times for $N > 2400$ can be attributed to the ''mean field'' nature of the calculation of $\omega_3$, i.e., to our assumption that (for a given $N$) the detachment of each motor head leads to the same energy gain. In reality, the energy change upon detachment of a motor depends, in some complex manner, on a number of factors such as the positions and chemical states of the motors. Motors which release higher energy will detach at higher rates, and the detachment of these ''energetic'' motors will lead to the release of much of the elastic energy stored in the actin track. We, therefore, conclude that within the mean field approach, the number of disconnecting motors and the frequency of detachment events are probably over-estimated. This systematic error of the mean field calculation increases with $N$, and the result of this is the decrease of $\tau_{rev}$ in this regime, which is not observed experimentally. Recently, we have demonstrated using a detailed statistical-mechanical calculation that in some cases, perfectly a-polar filaments may undergo a biased bidirectional motion with a net drift (43). This interesting effect cannot be explained within the mean field picture presented here.

## 4. Discussion and summary

In recent years, there has been a growing interest in the collective behavior of molecular motors which is ubiquitous in biology and physiology. Much progress has been achieved experimentally using bio-mimetic systems with new assays, and through a variety of theoretical models which have been proposed to interpret the experimental results. In this review we have focused on cooperative *bidirectional* motion which is one of the more fascinating phenomena associated with collective motor behavior. Bidirectional ("back and forth") motion originates from the work of a group of motors that exert opposite forces on a filament. Changes in the direction of the motion occur as a result of the motors binding to and unbinding from the cytoskeleton track. Recently, we presented a novel motility assay for bidirectional motion. Our studies showed that the characteristic reversal times of the dynamics are independent of the number of motors interacting with the track. This was an unexpected result since the existing theoretical models [both the many-motor ratchet model (35) and the TOW model for large $N$ (23)] predict an exponential growth of $\tau_{rev}$ with $N$. This prediction originates from the absence of crosstalk between the motors and the lack of correlations between their attachment/detachment states. In other words, the only coupling between motors assumed in these models is their physical linkage through their backbone from one side and the cytoskeleton track from the other.

Do other types of coupling between the motors exist, that fundamentally change the nature of their collective behavior? Our work presents a surprising answer to this question. The motors crosstalk with each other through the elasticity of the track. The forces which they exert on the track lead to the build-up of a tensile stress in the filament which can be relaxed by reducing the number of attached motors. *The elastic crosstalk between the motors does not lead to correlations between the binding states of specific motors (at least within our mean-field picture), but it makes the detachments rates of the motors highly sensitive to the number of attached motors.* Our detailed calculations show that this indirect cooperativity effect eliminates the exponential dependence of the reversal times on the number of motors and, thus, largely explains the experimental measurements. Since the size of $\tau_{rev}$ is often taken as a measure for the degree of cooperativity (the more cooperative the motors are, the more persistent is the movement and the longer are the periods of unidirectional transport), we reach the somewhat surprising finding that the elasticity-mediated crosstalk negatively affects the degree of cooperativity between motors.

We thank our students David Gillo, Barak Gilboa, and Barak Gur who conducted much of the research presented here. We also thank Yariv Kafri and Haim Diamant for useful discussions. The work was supported by the Israel Science Foundation (grant No. 551/04).